\documentclass[useAMS,usenatbib,times]{mn2e}
\input{epsf.sty}

\def\msun {{\rm M}$_{\odot}$}

\def\lsim{\mathrel{\hbox{\rlap{\hbox{\lower4pt\hbox{$\sim$}}}\hbox{$<$}}}}
\def\etal  {\rm {et al.}\rm}
\def\rmd {{\rm d}}

\begin{document}

\title[Clustering of Bright Lyman Break Galaxies in the ODT Survey]
{The Oxford-Dartmouth Thirty Degree Survey II: Clustering of Bright Lyman Break Galaxies - Strong Luminosity Dependent Bias at z=4} 
\author[Allen \etal]
{
Paul D. Allen$^{1,2}$\thanks{paul@mso.anu.edu.au}, Leonidas A. Moustakas$^{1,3}$, Gavin Dalton$^{1,4}$, Emily MacDonald$^{1}$
\newauthor
Chris Blake $^{1,5}$, Lee Clewley$^{1}$, Catherine Heymans$^{1,6}$ \& Gary Wegner$^{7}$
\\
$^{1}$Astrophysics, University of Oxford, Keble Road, Oxford OX1 3RH, UK. \\
$^{2}$RSAA, The Australian National University, Mount Stromlo Observatory, Cotter Rd, Weston, ACT 2611, Australia.\\
$^{3}$Space Telescope Science Institute, 3700 San Martin Drive, Baltimore, MD 21218, USA.\\
$^{4}$Space Science and Technology Division, Rutherford Appleton Lab., Didcot, 
OX11 OQX, UK.\\
$^{5}$School of Physics, University of New South Wales, Sydney, NSW 2052, Australia.\\
$^{6}$Max-Planck-Institut f\"{u}r Astronomie, K\"{o}nigstuhl 17, D-69117, Heidelberg, Germany.\\ 
$^{7}$Department of Physics and Astronomy, Dartmouth College, 6127 Wilder Laboratory, Hanover, NH 03755, USA.\\ 
}

\maketitle 

\begin{abstract}
We present measurements of the clustering properties of
bright ($L>L_{*}$) z$\sim$4 Lyman Break Galaxies (LBGs) selected from the
Oxford-Dartmouth Thirty Degree Survey (ODT). We describe techniques
used to select and evaluate our candidates and calculate the angular
correlation function which we find best fitted by a power law,
$\omega(\theta)=A_{w}\theta^{-\beta}$ with $A_{w}=15.4$ (with $\theta$ 
in arcseconds), using a constrained slope of $\beta=0.8$. Using a redshift 
distribution consistent with photometric models, we deproject this correlation 
function and find a comoving $r_{0}=11.4_{-1.9}^{+1.7}$\,h$_{100}^{-1}$\,Mpc 
in a $\Omega_m=0.3$ flat $\Lambda$ cosmology for $i_{AB}\leq24.5$. 
This corresponds to a linear bias value of $b=8.1_{-2.6}^{+2.0}$ (assuming 
$\sigma_{8}=0.9$). These data show a significantly larger 
$r_{0}$ and $b$ than previous studies at $z\sim4$. We interpret this as
evidence that the brightest LBGs have a larger bias than fainter ones, 
indicating a strong luminosity dependence for the measured bias of an LBG 
sample. Comparing this against recent results in the literature at fainter 
(sub-$L_{*}$) limiting magnitudes, and with simple models describing 
the relationship between LBGs and dark matter haloes, we discuss the 
implications on the implied environments and nature of LBGs. It seems that the 
brightest LBGs (in contrast with the majority sub-$L_{*}$ population), 
have clustering properties, and host dark matter halo masses, that are 
consistent with them being progenitors of the most massive galaxies today.
\end{abstract}


\section{Introduction}
\label{sec:intro}

The study of the universe at very high redshifts has expanded rapidly
over the last decade. It is now possible to observe galaxies over more than
90\% of the age of the Universe. One of the most significant
breakthroughs has been the discovery of a population of
strongly-clustered, star forming galaxies at $2.5<z<4.5$, using the
Lyman-break technique pioneered by \citet{s_and_h}.
It is possible to select significant numbers of these galaxies using
deep ground based multi-colour imaging \citep[e.g.,][]{steidel96}.

Lyman Break Galaxies (LBGs) represent the largest known population of
high redshift objects, and therefore present a window into an
important stage in the formation of galaxies and large-scale
structure. Much is now known about their physical characteristics
\citep{pettini01,shapley,shapley03}. They are somewhat dusty starburst
galaxies with star formation rates in the range 10- few 100s \msun
yr$^{-1}$ contributing a highly significant fraction of the stars
formed at $z\sim2.5$--$5$ \citep{as}.

Comparison of their clustering properties and number densities with
semi-analytic models suggests that they are either relatively small
galaxies, experiencing brief and infrequent bursts of star formation
that are primarily driven by galaxy-galaxy mergers (`The Collisional
Starburst Model'; \citealt*{spf}), or in very massive environments with
large reservoirs of gas, becoming massive $L_{*}$ galaxies today
(`The Massive Halo Model'; \citealt{steidel96,baugh}).

Although these scenarios have been tested against models of galaxy
formation using the LBG angular correlation function on different
scales, and as a function of luminosity \citep*{bullock02,wechsler01}
the results are still somewhat dependent on the observational sample
used. Correlation scale measurements range from r$_{0} =
2$--$12\,h_{100}^{-1}$\,Mpc
\citep{adelberger98,giav98,steidel98,arnouts,g_and_d01,ouchi,adelberger03,foucaud}. 
The range is likely due in part to cosmic variance because of relatively small 
sample sizes and may also reflect the luminosity dependence of clustering and 
the effects of small scale clustering.  

The data, (especially in the more studied $z\sim3$ population) do not yet 
provide unequivocal answers. 
\citet{p_and_g02} find a lack of power in the
angular correlation function on small scales suggesting few LBG close
pairs. \citet{g_and_d01} and \citet{foucaud} find fainter LBGs less strongly 
clustered than brighter ones. Conversely, the clustering results at $z\sim4$ 
of \citet{ouchi} imply an excess of LBG close pairs from which they
estimate an LBG merger rate. Using $z\sim3$ clustering data provided by 
Adelberger et al., several authors \citep{spf,wechsler01,bullock02} have 
shown that the collisional starburst type models are marginally more 
favourable than massive halo models. However it seems clear that current 
data sets are not extensive enough, and do not have accurate enough photometric
redshifts in large samples, to provide very strong constraints on the
relationship between LBGs and their host dark matter haloes. 
To begin to explore these questions it is useful to have very wide and 
deep multi-band imaging, to optimise the selection and redshift determination 
for many thousands of LBG candidates, over a range in absolute magnitudes.
The ODT is one such survey.

The structure of this paper is as follows.
In \S\,\ref{sec:odt} we present the ODT survey, and its salient 
characteristics.  The candidate
selection techniques, and their limitations, are discussed in
\S\,\ref{sec:candidate}.  The general statistical properties of
$z\sim4$ LBGs are presented in \S\,\ref{sec:genprops}, and the
projected and spatial correlation functions are the topic of
\S\,\ref{sec:corr}.  
These results are placed in a theoretical context,
focusing on the nature of the environments in which LBGs exist
in \S\,\ref{sec:disc}, the main conclusions finally being
drawn in \S\,\ref{sec:conc}.  Throughout this paper we use AB
magnitudes, and assume a ($\Omega_{m}$,$\Omega_{\Lambda}$,$\sigma_{8}$) =
($0.3,0.7,0.9$) cosmology, unless otherwise stated, with
$H_0=100\,h_{100}$\,km\,s$^{-1}$\,Mpc$^{-1}$.

\section{The Oxford-Dartmouth Thirty Degree Survey}\label{sec:odt}

\begin{figure}
\begin{center}
{\leavevmode \epsfxsize=8.0cm \epsfysize=4.2cm \epsfbox{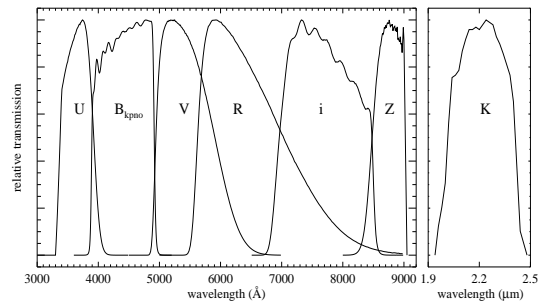}} 
\end{center}
\caption{Filter curves for the full set of filters employed in the ODT
Survey, convolved with the chip QE curve.}
\label{fig:filters}
\end{figure}

The Oxford-Dartmouth Thirty Degree Survey (ODT) is a deep-wide survey
using the Wide Field Camera  (WFC) on the 2.5-m Isaac Newton Telescope (INT) 
at La Palma. The survey employs six broad-band filters in $UBVRi'Z$, and the 
transmission curves are provided in Figure ~\ref{fig:filters}. When completed 
the survey will cover $\sim$30 deg$^2$ in $BVRi'Z$ to
$R_{5\sigma}=25.25$ in four fields of $5-10$\,deg$^2$ each. The coordinates
of these four fields are provided in Table~\ref{tab:fields}. Presently,
approximately $25$\,deg$^2$ of the survey have been observed in
$BVRi'Z$, although only data from the Andromeda field are analysed
in detail here. The $U$ data has only been observed in the best conditions,
and currently covers $\sim$1 deg$^2$, all of which is in the Andromeda 
field. We are also currently undertaking a $K$-band survey to
$K_{5\sigma}\approx18.5$ using the 1.3-m McGraw-Hill Telescope at the
MDM Observatory on Kitt Peak \citep{olding}. The three largest fields are also 
covered at radio frequencies with the VLA in $A$ and $D$ array at 1.4GHz,
and the Lynx field is covered by low frequency radio observations with the 
VLA in $A$ array at 74MHz and 330MHz.
In addition, we are able to obtain redshifts for several sources in the survey 
which are part of the Texas-Oxford One-Thousand (TOOT) redshift survey of
radio sources \citep{toot}.

Whilst the ODT survey has a number of scientific aims, one of the
principal projects is the detection of large numbers of LBGs (at
$z\sim$4) over a wider field than previous studies. The
ODT survey reaches depths comparable to previous studies of LBGs (e.g
\citealt{steidel99,ouchi}) but over a significantly larger area of sky 
(c.f. $\sim$900 square arcmin in \citealt{ouchi}). The ODT therefore 
provides an ideal data set for the selection of a sample of the 
\emph{brightest} LBGs. 

\begin{table}
\begin{center}
\begin{tabular}{ccccc} \\
\hline\\
Field & $\alpha$(J2000) & $\delta$(J2000) & $l$ &$b$\\\hline
Andromeda$^{*}$ & 00 18 24 & +34 52 00 & 115 & -27\\
Lynx$^{*\dagger}$ & 09 09 45 & +40 50 00 & 181 & +42\\
Hercules$^{*}$ & 16 39 30 & +45 24 00 & 70 & +41\\
Virgo & 13 40 00 & +02 30 00 & 330 & +62\\\hline\\ 
\end{tabular}
\caption{$3^\circ\times 3^\circ$ field centres
for the ODT survey in equatorial and Galactic coordinates. $^{*}$Fields with
1.4GHz radio data. $^{\dagger}$Field with 74MHz and 330MHz radio data.}
\label{tab:fields}
\end{center}
\end{table}

The fields discussed in this paper were observed with the 2.5-m INT
during August 1998 and September 2000. Approximate 5$\sigma$ (isophotal) 
depths for the data along with exposure times are shown in Table
\ref{tab:depth}. The median seeing for these observations is
$\sim$1.1$^{\prime\prime}$.

\begin{table}
\begin{center}
\begin{tabular}{ccc} \\
\hline\\
Filter & Exposure &5-$\sigma$ \\
       &  Time  & Limit   \\\hline
$U$ & 6$\times$1200$s$ & 26.0 \\
$B$ & 3$\times$900$s$ & 26.0 \\
$V$ & 3$\times$1000$s$ & 25.5 \\
$R$ & 3$\times$1200$s$ & 25.3 \\
$i'$ & 3$\times$1100$s$ & 24.5 \\
$Z$ & 1$\times$600$s$ & 22.5 \\
\hline\\
\end{tabular}
\caption{Exposure times and detection limits (AB at 5$\sigma$
isophotal) for the different passbands used for the ODT survey. With
the exception of the $U$ images (which use 6), each pointing is broken
up into three separate exposures with $5^{\prime\prime}$ offsets of
the telescope. The detection limits are approximately equal to 5$\sigma$
detections using a 2\arcsec circular aperture and 3$\sigma$ detections with a
3.3\arcsec circular aperture.}
\label{tab:depth}
\end{center}	
\end{table}

The reduction of CCD frames was carried out using the {\tt IRAF}
package and photometry calibrated for each frame by observing fields
of \citet{landolt} on photometric nights. The thinned chips in the INT
WFC mean that fringing of sky lines becomes a problem in
$i'$, $Z$, and to a lesser extent, $R$ data. Fringing was removed
from each image in turn by using a fringe frame generated by the
combination of individual images for each chip. This method was successful
for the $i'$ and $R-$band data, but the $Z-$band data have more significant
fringing, and little of the acquired data have been incorporated to date.
Image detections were carried out using the SExtractor package \citep{bertin}, 
although we chose to use our own background subtraction method, due to 
significant background gradients present in some of our images.

Each WFC pointing was arranged in a diagonal grid to
cover each field, with several overlaps between adjacent
pointings. This allows objects in overlaps to be matched and a common
photometric zeropoint to be applied to each field, using the method of
\citet{gbrook}. This method was used to obtain a common zeropoint for the 
$V$-band data which has the smallest scatter between chips and is not afflicted
by fringing from sky lines.

In order to obtain consistent colours throughout the survey, photometry for
other bands was corrected relative to the V-band data. This was done  
by minimising the deviation of the colours of ODT stellar objects from the 
colours of stars from \citet{pickles} in the colour-colour plane. The ODT
survey and our data reduction process is described in more detail in 
\citet{macd}.

\section{Candidate Selection}\label{sec:candidate} 

The LBG population of galaxies are traditionally selected by
exploiting the break in their spectra shortwards of the Lyman-Limit
($\lambda_{rest}=912$\,\AA). The break is caused by an intrinsic drop
in the spectra of the massive stars which dominate the LBG spectrum,
and more importantly, the absorption of photons shortwards of the
Lyman-Limit by H\,{\scriptsize I} in the interstellar medium of the
LBG. The break is further accentuated by absorption from Lyman-Limit
Systems along the line-of-sight to the galaxy. By choosing appropriate
filter systems to straddle this break, it becomes possible to select
high-$z$ galaxy candidates.
	
The `standard' method of LBG selection uses a 3-band filter set to
isolate high-$z$ galaxies in the colour-colour plane
(e.g. \citealt{steidel99,ouchi}).  The selection region is defined
using the predicted colours of synthetic galaxy spectra at high
redshifts, and where spectroscopy is available, this can often be
refined. However, for many surveys such as the ODT or the Hubble Deep Field
(HDF), further information from extra filters may be available. By using this 
extra information we can obtain photometric redshifts for objects in a
survey (e.g. \citealt{arnouts,arnouts02}), and select the high redshift
objects. It is also possible, with multi-band data, to consider
more than one filter combination as we discuss in the next section.

\subsection{Models of High-$z$ Galaxy Colours}
\label{sec:model}

\begin{figure}
\begin{center}
{\leavevmode \epsfxsize=8.0cm \epsfysize=8.0cm \epsfbox{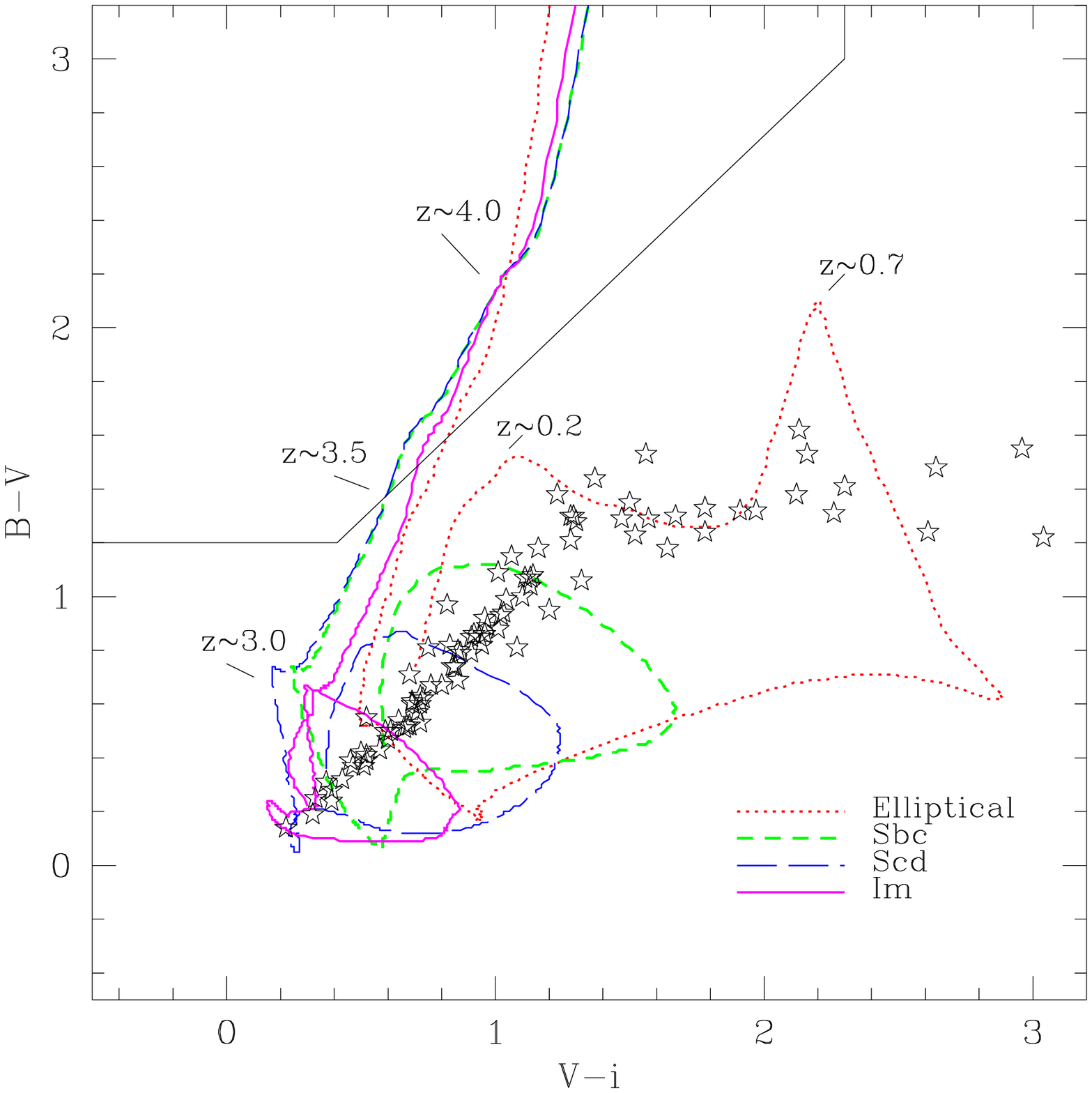}} 
\end{center}
\caption{
Model $B-V$ versus $V-i'$ colours for galaxies to $z\sim5$. Four
SEDs from \citet{cww} are used. No evolution is assumed, but the SEDs are 
modified to account for the affects of attenuation due to intergalactic 
absorbers at high redshift \citep{madau}. The four SED types of \citet{cww}
are shown as; a red dotted line (type E), a green short-broken line (type Sbc),
a blue long-broken line (type Scd), and a magenta solid line (type Im). Also
shown as black open stars, are the colours of Galactic stars, using the 
stellar library of \citet{pickles}. Important redshifts are labelled, most 
importantly, galaxies with $z>3.5$ (i.e. LBGs), and, on the Elliptical track,
intermediate redshift ($z=0.2-1.1$) objects, which are a prime source of 
contamination in a high redshift sample (see Section \ref{sec:complete}). The 
final ODT selection window for selecting $z\sim4$ objects is bounded by the 
solid black line.}
\label{fig:bvvi_model}
\end{figure}

\begin{figure}
\begin{center}
{\leavevmode \epsfxsize=8.0cm \epsfysize=8.0cm \epsfbox{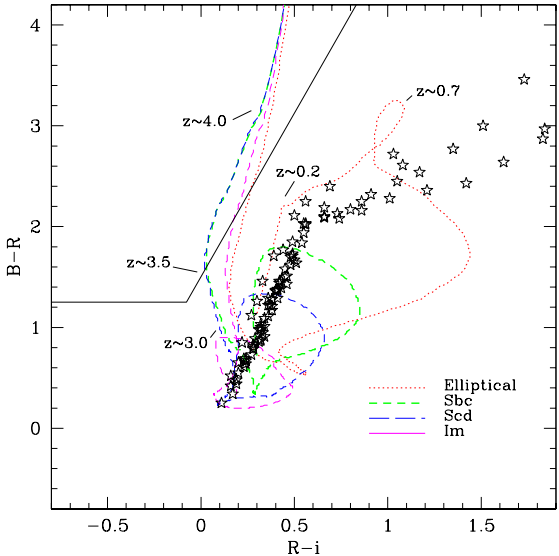}} 
\end{center}
\caption{As Figure~\ref{fig:bvvi_model} but using model $B-R$ versus $R-i'$ 
colours.}
\label{fig:brri_model}
\end{figure}

In order to define our selection criteria for LBG candidates we plot
colours for model galaxies (see Figures~\ref{fig:bvvi_model} 
and~\ref{fig:brri_model}) using
extended \citet*{cww} empirical spectral templates convolved with the
INT WFC filter set. We consider two different regimes;
selection using $B-V/V-i'$ colours, and selection with $B-R/R-i'$ colours.
In addition, we plot stellar colours
using the stellar SED library of \citet{pickles}. Galaxy colours are also 
adjusted for absorption from the IGM using the method of \citet{madau}.
This attenuation of flux due to the opacity of the IGM proves to be a
significant contributor to the final observed colours of high-$z$ objects. 
In addition to these empirical templates, we also compare with the 
model grids of \citet{stevens}, which are based on the 
Pegase models of \citet{pegase} and the standard stars of 
\citet{landolt}, again using our INT filter set. These models contain SEDs with
a range of starburst and star formation histories along with different amounts 
of dust. However they compare favourably with the simple model colours shown
in Figures \ref{fig:bvvi_model} and \ref{fig:brri_model}. The colours of 
$z\sim4$ objects are dominated by the effects of the Lyman break, and LBGs 
occupy essentially the same region of colour space in both models.

In order to efficiently select high redshift objects, we define the
following colour-colour cuts based upon the model galaxy colours shown
in Figure~\ref{fig:bvvi_model} for $B-V/V-i$ selection.

\noindent For $-1.0<(V-i')<0.41$, 
\begin{equation}
(B-V) > 1.2.  
\end{equation}
For $0.41 < (V-i') < 2.3$, 
\begin{equation}
(B-V) > 0.95\times(V-\emph{i'}) + 0.81. 
\end{equation}

\noindent
Objects with $(V-i')<-1.0$ and $(V-i')>2.3$ are excluded. 
For $B-R/R-i$ selection, we use the following colour cuts based on 
Figure~\ref{fig:brri_model}.
  
\noindent For $-1.0<(R-i')<-0.07$, 
\begin{equation}
(B-R) > 1.25.  
\end{equation}
For $-0.07 < (R-i') < 0.9$, 
\begin{equation}
(B-R) > 3.25\times(R-\emph{i'}) + 1.5. 
\end{equation}

\noindent
Objects with $(R-i')<-1.0$ and $(R-i')>0.9$ are excluded.

\subsection{Colour-Colour Selection}
\label{sec:colcol}

\begin{figure}
\begin{center}
{\leavevmode \epsfxsize=8.0cm \epsfysize=8.0cm \epsfbox{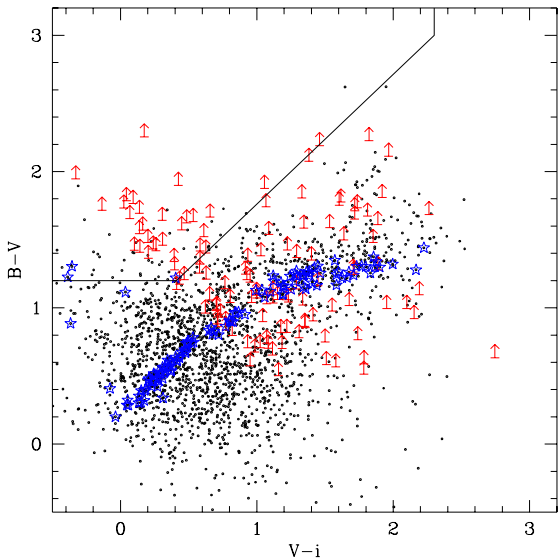}}
\end{center}
\caption{$B-V$ versus $V-i'$ colours for ODT data. For clarity, only 5000
randomly chosen objects are shown. The resulting colour distribution is
representative of the full data set.
All objects with detections in $V$, $R$, and $i'$ down to
the limits of the survey (Table \ref{tab:depth}) are shown.
Objects with no B detection are shown as red arrows. 
Stellar objects are designated by blue open stars, these were selected 
using SExtractor's stellarity parameter, and objects with 
stellarity $>$ 0.8 and $R<23.0$ are plotted. All other objects
are shown as black circles. The stellar locus is clearly visible and
this provides a useful cross-check of the photometric consistency
between fields. The selection region determined from the models
discussed in section \ref{sec:model} is bounded by the black solid
line.}
\label{fig:bvvi}
\end{figure}

\begin{figure}
\begin{center}
{\leavevmode \epsfxsize=8.0cm \epsfysize=8.0cm \epsfbox{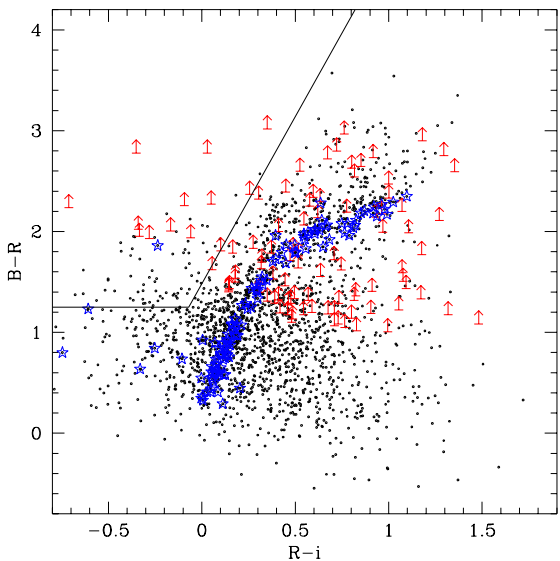}}
\end{center}
\caption{As Figure~\ref{fig:bvvi} but using the $B-R$ versus $R-i'$ colours 
for a $5000$ object subset from the ODT.}
\label{fig:brri}
\end{figure}

Colours for a small subsection of the ODT survey are shown in Figures
\ref{fig:bvvi} and \ref{fig:brri} along with the LBG selection region in 
colour-colour space.  We also require a detection in the $V$, $R$ and $i'$ 
bands, increasing the probability of only detecting real objects close to 
the faint limits of the survey. In addition, only candidates fainter than
$i'=23.0$ are considered in order to reduce contamination from lower
redshift ellipticals with similar colours to LBGs \citep{steidel99}. Finally, 
candidates that have $B>25.5$ or are not detected by SExtractor in $B$, are
treated as non-detections and are given an limiting magnitude of $B=25.5$.  

Given that we have 4-band information available to us, we impose the 
additional criterion that the object lies in the selection regions of both
$B-V/V-i'$ and $B-R/R-i'$ (Figures \ref{fig:bvvi} and \ref{fig:brri}). 
Since the surface densities of objects are much higher just outside the 
selection region than just inside, 
low redshift objects are much more likely to be scattered into the selection
region than vice-versa. This is somewhat reduced by requiring that an object
is selected in both planes, and indeed
we obtain candidate surface densities comparable to other searches for $z\sim4$
objects (see Figure~\ref{fig:sd}). The contamination of the sample, and the 
effects of this restriction are discussed further in the next section.

\subsection{Contamination}
\label{sec:complete}
 
Although our selection criteria are fairly conservative, contamination of 
a non-spectroscopic sample such of this is expected. The prime candidates
for contamination are foreground Galactic stars, high redshift quasars, and
most importantly, elliptical galaxies with $0.2<z<1.1$. Spectroscopic
follow-up of a sample selected using similar selection criteria by
\citet{steidel99} confirmed that about 20$\%$ of candidates were in fact
elliptical galaxies at intermediate redshifts.
The peak of the luminosity function for elliptical galaxies with $z\sim0.7$ 
corresponds to galaxies with $i'\sim22.0$ \citep{bell}. Having a bright 
cut-off of $i'=23.0$ for the sample should reduce contamination from these 
objects significantly, although a number of faint potential contaminants are 
expected to remain. In the next Section, we consider two faint limits of 
$i'=24.0$ and $i'=24.5$, and so estimate contamination levels for both limits
here.

In Figures \ref{fig:bvvi_model} and \ref{fig:brri_model}, the
elliptical track is noticeably more extended than for other types. Objects
in the range $0.2<z<1.1$ are both close enough, and faint enough, to scatter
into the LBG selection region. In order to estimate the potential level
of contamination from these objects, we simulate their expected colours by 
combining their colours from an SED with typical photometric errors. 
We then normalise their numbers with an appropriate luminosity function, and
test how many objects would meet our selection criteria. To obtain the 
colours of elliptical galaxies we use the SED from \citet{cww}, as 
plotted in Figures \ref{fig:bvvi_model} and \ref{fig:brri_model}. The expected
number of galaxies can then be obtained by integrating the galaxy luminosity
function using \citep{gardner}:

\begin{equation}
n(m,z)\,\rmd m\rmd z=\frac{\Omega}{4\pi}\frac{\rmd V}{\rmd z}\phi(m,z)\,\rmd m\rmd z,
\end{equation}

\noindent
where $\phi(m,z)$ is the galaxy luminosity function, and $\rmd V/\rmd z$ is the
comoving volume element at redshift $z$. We use the $R$-band luminosity 
function for elliptical galaxies from the ESO-Sculptor survey at $z\sim0.5$ 
\citep{sculptor}. The rest-frame $R-$band and redshift range covered by
this luminosity function compares favourably to the objects of interest here. 
Little, or no evolution of the luminosity function is expected in our redshift
range of interest \citep{bell}. To simulate photometric errors, a random 
error is generated using a probability distribution describing 
the typical error as a function of magnitude. 
The error in each band is then added to the respective magnitude.
If just one combination of colours is used (especially $B-V/V-i'$), then the 
contamination can be quite considerable (up to $27\%$, see Table 
\ref{tab:contam}). However, by requiring that LBGs meet both selection 
criteria ($B-V/V-i'$ and $B-R/R-i'$), the expected contamination from 
intermediate redshift galaxies can be reduced significantly. For $i'<24.0$ we 
find a contamination of $4\%$, and for $i'<24.5$ we obtain $3\%$.

The ODT Andromeda field (i.e. the data discussed in this paper) has a fairly
low Galactic latitude of $l=-27^{\circ}$, and therefore Galactic stars must be
considered as potential contaminants. At the magnitudes considered here, the 
SExtractor star-galaxy classifier (in general) fails to distinguish cleanly 
between stellar and galactic light profiles. In particular, red stars 
are prone to be found in the selection window due to extreme intrinsic 
colours, or because they lie close to the selection window and are scattered in
because of photometric errors, or a combination of both.
In order to estimate the contamination due to stars, a model of the Galaxy 
\citep{robin_creze,robin96} is used to generate a mock 
catalogue\footnote{http://www.obs-besancon.fr/www/modele/} of Galactic 
stars at the Galactic latitude and longitude of the ODT Andromeda field. 
This model contains elements from the thin and thick disk, along with the 
spheroid of the Galaxy, and a simple model of Galactic extinction 
\citep{robin_creze}. 
The $V-$band number counts generated by this model were 
compared with the `Bahcall-Soneira' model \citep{bahcall}, and with stellar 
number counts over the range $18<V<22$ from the data itself, and found to be
consistent.
\begin{table*}
\begin{center}
\begin{tabular}{ccccccccccccc}\\
\hline\\
Magnitude Range && \multicolumn{3}{c}{Stars} && \multicolumn{3}{c}{Ellipticals} && \multicolumn{3}{c}{Total}\\
\cline{3-5}\cline{7-9}\cline{11-13}\\
$i'$ && $BVi'$ & $BRi'$ & Both && $BVi'$ & $BRi'$ & Both && $BVi'$ & $BRi'$ & Both \\
\hline
$23.0-24.0$ && $22\%$ & $30\%$ &$8\%$ &&$27\%$ &$4\%$ & $4\%$&& $49\%$ & $34\%$ & {{\bf 12$\%$}}\\
$23.0-24.5$ && $20\%$ & $28\%$ &$5\%$ &&$15\%$ &$7\%$ & $3\%$&& $35\%$ & $35\%$ & {{\bf 8$\%$}}\\
\hline\\ 
\end{tabular}
\caption{Summary of expected contamination for stars and elliptical galaxies.
The contamination fraction can be significantly reduced by using both selection
criteria.}
\label{tab:contam}
\end{center}
\end{table*}

\begin{table*}
\begin{center}
\begin{tabular}{cccccccccc} \\
\hline\\
Field$^{a}$ & RA$^{b}$ & Dec$^{c}$ & Seeing$^{d}$ & Area$^{e}$ & $N_{{{\rm LBG}}}^{f}$ & $B$ & $V$ & $R$& $i'$\\
\hline
I012$^{*}$& 00 11 38.47 & +35 39 36.7 & 0.95 & 675.7 & 22 & 25.8 & 25.1 & 25.1 & 24.7\\
I013 & 00 13 49.66 & +35 45 47.8 & 1.07 & 460.7 & 9 & 25.7 & 25.1 & 24.6 & 24.4\\
I015& 00 16 02.41 & +35 51 51.3 & 1.24 & 424.7 & 10 & 25.5 & 25.0 & 25.0 & 24.5\\
I026& 00 10 29.80 & +35 03 39.3 & 1.09 & 538.7 & 18 & 25.5 & 24.6 & 24.8 & 24.5\\
I029& 00 18 19.46 & +35 34 47.4 & 1.16 & 458.8 & 15 & 25.5 & 25.1 & 25.0 & 24.4\\
\hline\\
\end{tabular}
\caption{Data for ODT Andromeda fields used in this paper. These
fields form a single contiguous field. (a) Field name. (b) RA of field 
centre. (hms) (c) Dec of field centre. (dms) (d) seeing (averaged over $BVRi'$ 
bands). (e) Area in square arcminutes (after removal of bad areas of 
the chip, bright stars, satellite trails etc.). (f) Number of LBGs with $i'<24.0$. The final four columns give the magnitude at which the total number counts turn over in each field. $^{*}$Field used for the $i<24.5$ LBG sample.}
\label{tab:pointings}
\end{center}	
\end{table*}

The model produces (without photometric errors) the expected mixture of stars
at given Galactic coordinates in Johnson-Cousins $UBVR_{c}I_{c}$ (Vega) 
magnitudes. These were then transformed to the ODT $UBVRi'$ (AB) system and 
combined with a function to simulate a typical photometric error in each 
band, as done for the simulated galaxies. 
Our selection criteria were then applied to these 
stars. In total, 17 stars (in one square degree) are found to match the 
selection criteria to $i'=24.5$, 7 of which have $i'<24.0$. Given the surface 
densities of LBG candidates (see Section \ref{sec:sd}), this corresponds to a 
contamination of $8\%$ for $i'<24.0$, and $5\%$ for $i'<24.5$. We note that, 
again, these results increase significantly if only $B-V/V-i'$ or $B-R/R-i'$ 
selection is used (see Table \ref{tab:contam}).

Finally, using the quasar luminosity function at $z\sim4$ \citep{fan}, we 
expect that even if all high-$z$ quasars were to have colours consistent with 
LBGs, their expected number densities make any contamination from this 
population negligible. Our final contamination levels are therefore $12\%$ for
$i'<24.0$, and $8\%$ for $i'<24.5$. The contamination results are summarised 
in Table \ref{tab:contam}. These results compare favourably with 
other LBG surveys \citep[e.g.][]{ouchi}, although the overall contamination 
level is reduced through the advantage of having two sets of selection 
criteria. Objects are less likely to be scattered into both selection windows
through photometric errors than just one.
We also note that our {{\it observed}} surface densities are fully consistent 
with other surveys (see Section \ref{sec:sd}), implying similar levels of 
contamination and completeness.

\subsection{The ODT LBG Samples}

This paper uses the first 2 square degrees of ODT data in the Andromeda field,
which was the first part of the ODT survey to be reduced and analysed.
Unfortunately the quality and depth of the survey is variable, and we 
restrict the data used in this paper to those regions with the best seeing.
In order to reduce contamination from any spurious signal caused by 
field-to-field variations we consider two samples of data, one to $i'<24.5$, 
and the other to $i'<24.0$. These were chosen to provide a 
contiguous area covered by all colours to the required depth, 
where we are confident of consistent completeness in all bands. When fields
overlap, the overlap regions are split equally (in R.A.) between the two 
fields, and objects in the region will have photometry measured from one of 
the fields.

\begin{table}
\begin{center}
\begin{tabular}{ccccc} \\
\hline\\
Sample & Limit ($i'$) & Fields & Number & Total Area\\
\hline
Bright & 24.0 & All & 74 & 0.81 deg$^{2}$\\
Faint & 24.5 &I012 & 66 & 676 arcmin$^{2}$\\
\hline\\
\end{tabular}
\caption{Summary of `bright' and `faint' LBG samples selected from the ODT.}
\label{tab:summ}
\end{center}	
\end{table}

\begin{figure}
\begin{center}
{\leavevmode \epsfxsize=8.0cm \epsfysize=8.0cm \epsfbox[28 90 592 654]{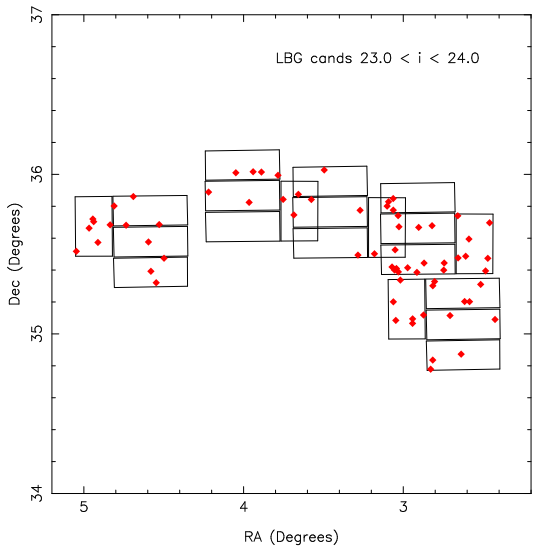}}
\end{center}
\caption{The distribution of our `bright' sample of LBG candidates, with $23<i'<24$. The outlines of the chips which make up the INT wide field camera pointings for this mosaic are shown.}
\label{fig:lbgs23_24}
\end{figure}

\begin{figure}
\begin{center}
{\leavevmode \epsfxsize=8.0cm \epsfysize=8.0cm \epsfbox[28 90 592 654]{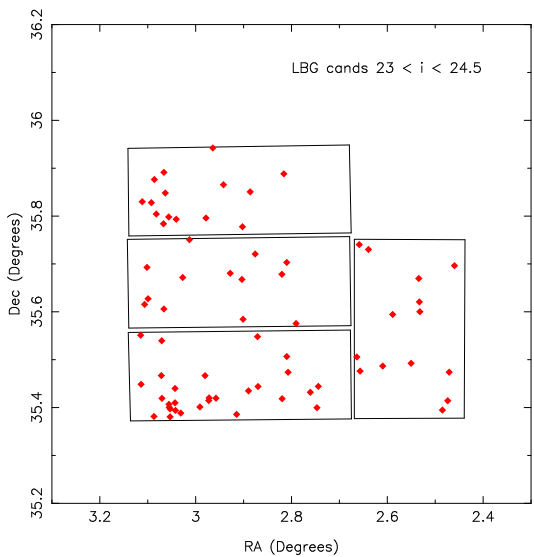}}
\end{center}
\caption{The distribution of our `faint' sample of LBG candidates, with $23<i'<24.5$. The outlines of the 4 chips which make up the INT wide field camera are shown.}
\label{fig:lbgs23_245}
\end{figure}

The properties of the fields used (including depths) are summarised in Table 
\ref{tab:pointings}, and Table \ref{tab:summ} summarises the properties of the
two selected samples. All five pointings are used in the bright ($i'<24.0$)
sample. These cover a region with good seeing and a
completeness beyond the required $i'<24.0$. The distribution of these objects 
is shown in Figure \ref{fig:lbgs23_24}. After removal of satellite trails, 
overlaps, diffraction spikes, and bright objects, the total area covered is
0.83 deg$^{2}$. This sample consists of 74 objects.
The `faint' ($i'<24.5$) sample, consists of just one INT WFC 
pointing covering 676 arcmin$^{2}$ (Field I012 in Table \ref{tab:pointings}). 
The distribution of these objects is shown in Figure \ref{fig:lbgs23_245}. This
sample is made up of 66 objects.

\section{General Properties of LBGs}\label{sec:genprops}

\subsection{Redshifts}
\label{subsec:photz}

Although the primary selection criterion for LBG candidates is the
effect of the integrated opacity by intergalactic H\,{\scriptsize I},
redshifts are an important part of the analysis and interpretation of
such samples. Obtaining spectroscopic redshifts for more than a very
small fraction of our LBG candidates is not currently forthcoming, and 
represents a major challenge for current instrumentation, even on $8-$m 
class telescopes. 

However, in this paper, their most important use is in their statistical 
distribution, $N(z)$, and the corresponding selection function.  This is 
used for deprojecting the angular correlation function to retrieve the spatial 
correlation function $\xi(r)$, and for detailed studies of the effects of 
different dark matter halo occupation function parameterisations on the global
statistics, but not for the selection of LBG candidates.

\subsection{The Selection Function}
\label{sec:Nz}

The selection function, as represented by the redshift distribution,
is approximated by a simple analytic model based on the combined
results from the colour-colour diagrams in Figures \ref{fig:bvvi_model} and 
\ref{fig:brri_model}, and the models of \citet{stevens}.
The main use of our redshift distribution will be the deprojection of the 
angular correlation function to obtain r$_{0}$ (Section~\ref{sec:corr}), and we
model redshift distributions using:

\begin{equation}\label{eq:nz}
N(z) = \frac{1}{{\sqrt{2\pi\sigma_{z}^{2}}}}
\exp\left[{-\frac{(z-\bar{z})^{2}}{2\sigma_{z}^{2}}}\right], 
\end{equation}

\noindent
where $\bar{z}$ is the mean, and $\sigma_{z}$ is the standard deviation.

\begin{figure}
\begin{center}
{\leavevmode \epsfxsize=8.0cm \epsfysize=8.0cm \epsfbox{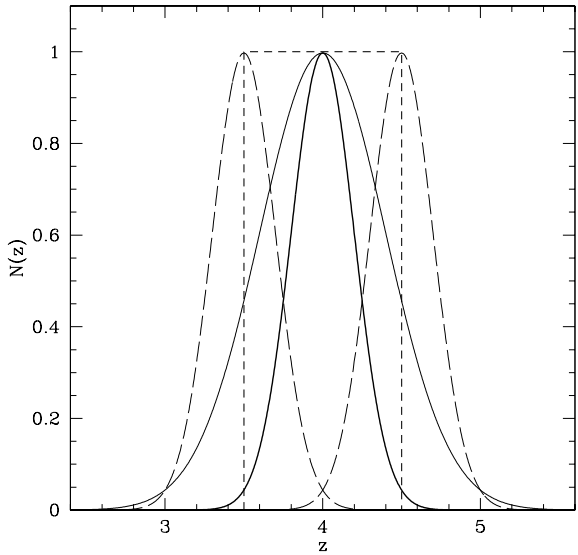}}
\end{center}
\caption{Normalised fiducial redshift distribution consistent with photometric 
models (bold solid line corresponding to a Gaussian with $\bar{z}=4.0$ and 
$\sigma_{z}=0.2$). The other lines show the range of selection functions
considered to test the sensitivity of the Limber deprojection to the assumed 
redshift distribution. The bold short-dashed line shows a `tophat' function
with centred on $\bar{z}=4.0$ with $\bar{z}\pm0.5$. The thin long 
dashed lines correspond to Gaussians with $\bar{z}=3.5$ and 4.5 with 
$\sigma_{z}$=0.2. Finally, the thin solid line corresponds to a much wider 
redshift distribution with $\sigma_{z}$=0.4.}
\label{fig:Nz}
\end{figure}

We find that $\bar{z}=4.0$ and $\sigma_{z}=0.2$ is a reasonable
fiducial set of values that reproduces the photometrically-determined
distributions. 
In order to test the dependence of $r_{0}$ on $N(z)$, we also consider other
selection functions with varying mean redshifts, $\bar{z}$, and dispersions,
$\sigma_{z}$, as well as modelling the redshift distribution as a simple
top-hat function. These selection functions are shown in Figure \ref{fig:Nz}.
However, as we discuss in Section~\ref{subsec:spatcorr}, at these redshifts,
the deprojected $r_{0}$ we obtain is quite insensitive to the $N(z)$ used.
We do not propagate systematic errors from the uncertainty in $N(z)$
in what follows.
  
\subsection{Surface and Space Densities}
\label{sec:sd}

\begin{figure}
\begin{center}
{\leavevmode \epsfxsize=8.0cm \epsfysize=8.0cm \epsfbox{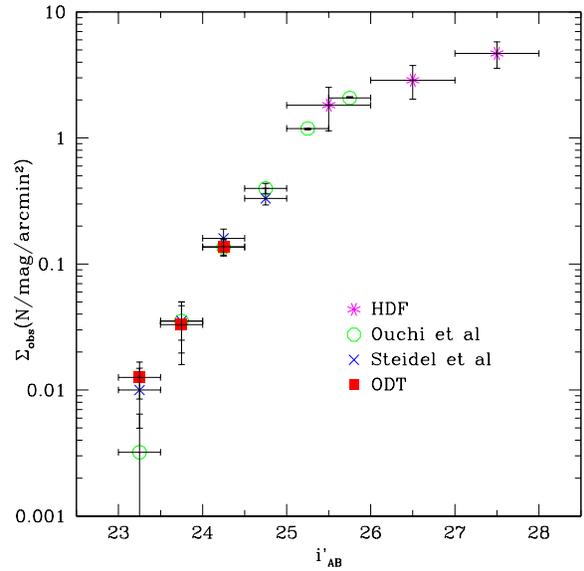}} 
\end{center}
\caption{The surface density of colour-colour selected LBGs in the ODT
Survey (filled squares). Also shown are data from \citet{ouchi} (open circles), \citet{steidel99} (crosses), and data from the Hubble Deep Field, \citet{arnouts} (asterisks). Surface densities are plotted per unit 
magnitude.}
\label{fig:sd}
\end{figure}

The surface density of objects is calculated by selecting
out the regions contaminated by bright stars, chip edge-effects, satellite
trails, exceptionally poor seeing and strong remnant fringe effects. The 
sample selection described here is performed in the $i'$-band, requiring 
detection in the $R$ and $V$ bands as well.  The measured surface densities, 
with Poisson uncertainties, are given in Table~\ref{tab:dens} and shown in
Figure~\ref{fig:sd}. In calculating these surface densities we only use 
those images that are both deep enough and have good seeing as
described in Section~\ref{sec:colcol}.  Figure~\ref{fig:sd} indicates 
that the ODT LBG surface densities compare well to other work 
\citep{steidel99,ouchi}. We also calculate the space density using: 

\begin{equation}
n_{\Sigma} = A_{\Omega} \Sigma \left[\int_0^{\infty} N(z) \frac{dV}{dz}\,dz\right]^{-1},
\end{equation}

\noindent
where $\Sigma$ is the measured surface density in arcmin$^{-2}$, as
shown in Figure~\ref{fig:sd}. We use $N(z)$ from Eq.~\ref{eq:nz}, and 
$dV(z)$ is the comoving differential volume element per square steradian
for the cosmology of choice.  The coefficient 
$A_{\Omega}=1.1818\times10^{7}$\,arcmin$^2$\,sr$^{-2}$.  The
calculated surface and space densities are presented in
Table~\ref{tab:dens}.  In the next Section we also explore how
sensitive the correlation function parameters are relative to 
the choice of $N(z)$, and to the limiting magnitude or chosen magnitude
range.  

\begin{table}
\begin{center}
\begin{tabular}{ccc} \\
\hline\\
magnitude range ($i'$) & $\Sigma$ (arcmin$^{-2}$) & $n_{\Sigma}$
($h_{100}^3$\,Mpc$^{-3}$) \\ 
\hline
23.0 -- 23.5 & $0.0065\pm0.0041$ & $6.29\pm5.30\times10^{-6}$ \\
23.5 -- 24.0 & $0.0165\pm0.0085$ & $1.60\pm1.10\times10^{-5}$ \\
24.0 -- 24.5 & $0.0685\pm0.0101$ & $6.63\pm1.31\times10^{-5}$ \\
\hline\\
\end{tabular}
\caption{The surface and space densities for different $i'$-band
selection magnitude ranges. Surface densities are per half magnitude bin.}
\label{tab:dens}
\end{center}	
\end{table}

\section{Correlation Functions}\label{sec:corr}

\subsection{Angular Correlation Function}
\label{subsec:angcor}

One of the more important results to come from LBG studies has been
the measure of their angular correlation function, and associated
$r_{0}$ and bias parameter, `$b$'. It has been shown
\citep[e.g][]{wechsler01,bullock02} that this can be used to constrain
the halo occupation function for LBGs (see Section \ref{sec:disc}), which 
in turn provides information on their typical masses and likely 
evolutionary fate \citep[e.g.][]{ms02}. 

The clustering of galaxies, as represented by the two-point correlation 
function, has been shown to be well approximated by a power
law \citep[e.g][]{peebles1980} of the form $\xi(r)=(r/r_{0})^{-\gamma}$. The 
angular projection of this also follows a power law of the form 
$w(\theta)=A_{w}\theta^{-\beta}$, where $\beta=\gamma-1$. 
We calculate the two-point angular correlation function,
$\omega$($\theta$), for $i'$-band selected galaxies, using the estimator 
of \citet{ls}:

\begin{equation}
\omega(\theta) = 1 +  \frac{DD(\theta)}{RR(\theta)}W_{1} 
                   - 2\frac{DR(\theta)}{RR(\theta)}W_{2}, 
\end{equation}

\noindent
where $DD(\theta)$, $RR(\theta)$, and $DR(\theta)$ are the numbers of
`data-data', `random-random' and `data-random' pairs, respectively,
and $W_{1}$ and $W_{2}$ account for the numbers of data and random
points ($N_{{{\rm ran}}}$ and $N_{{{\rm data}}}$) used to estimate the 
correlation function \citep{roche:99}. Here, 

\begin{equation}
W_{1}=\frac{N_{{{\rm ran}}}(N_{{{ \rm ran}}}-1)}{N_{{{\rm data}}}(N_{{{ \rm data}}}-1)},
\end{equation}
and 
\begin{equation}
W_{2}=\frac{(N_{{{\rm ran}}}-1)}{N_{{{\rm data}}}}. 
\end{equation}

\noindent
In the data we remove bright stars, diffraction spikes, bad columns,
etc., and constrain the random sample so that it follows the same
geometrical constraints as the data.  

The Poisson error for the angular correlation function can be calculated 
using:

\begin{equation}
\sigma_{\omega(\theta)} =
\frac{1+\omega(\theta)}{\sqrt{DD(\theta)}}.
\label{eq:wpoisson}  
\end{equation}

\noindent
However, $\sigma_{\omega(\theta)}$ is probably an underestimate of the actual 
error in $\omega(\theta)$ when the number of data points used is small 
\citep{baugh_error}. In order to obtain a more appropriate estimate of the 
errors on $w(\theta)$, we compute bootstrap errors \citep{ling86}. 

If a sample contains $n$ galaxies then a `bootstrap' sample can be created by
drawing (\emph{with replacement}), $n$ galaxies from the original galaxy sample.
This process can be repeated $N$ times and the correlation function, 
$w_{i}(\theta)$, calculated for each of the $N$ bootstrap samples (where 
$i=1,2,3...N$). The estimate of the error in $w(\theta)$ is then given by:

\begin{equation}
\sigma_{{{\rm boot}}}=\sqrt{\left(\sum_{i=1}^{N}\frac{[w_{i}(\theta)-w_{{{\rm av }}}(\theta)]^{2}}{N-1}\right)},
\end{equation}

\begin{figure}
\begin{center}
{\leavevmode \epsfxsize=8.0cm \epsfysize=8.0cm \epsfbox{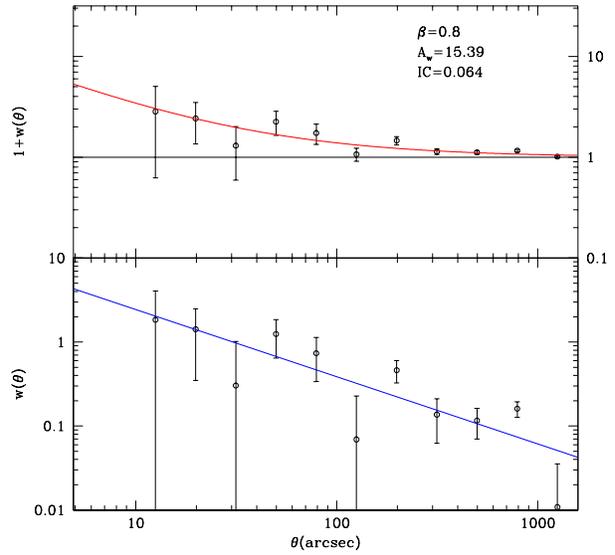}} 
\end{center}
\caption{The angular correlation function for 66 $i'<24.5$ LBGs (bottom 
panel). The fit is restricted to a fixed slope of $\beta=0.8$. The top 
panel shows $1+w(\theta)$. Data points are corrected for the 
integral constraint}
\label{fig:wtheta}
\end{figure}

\begin{figure}
\begin{center}
{\leavevmode \epsfxsize=8.0cm \epsfysize=8.0cm \epsfbox{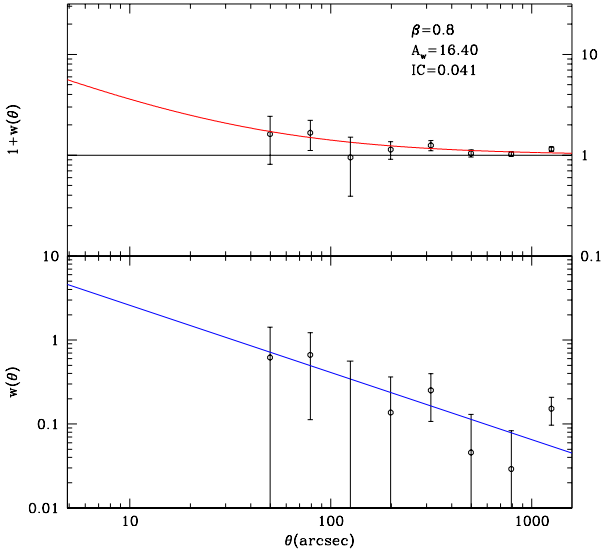}} 
\end{center}
\caption{The angular correlation function for 74 $i'<24.0$ LBGs (bottom 
panel). The fit is restricted to a fixed slope of $\beta=0.8$. The top panel
shows $1+w(\theta)$. Data points are corrected for the integral constraint.}
\label{fig:wtheta2}
\end{figure}

\noindent
The computed bootstrap errors are shown in Figures \ref{fig:wtheta} and 
\ref{fig:wtheta2}, and are used in what follows.

Due to the fact that we are
using a finite solid angle, we correct the estimate of
$\omega$($\theta$) for the integral constraint \citep{groth,roche:99}:

\begin{equation}
IC = \frac{1} {\Omega^{2}} \int\int \omega(\theta_{12})
\delta\Omega_{1} \delta\Omega_{2},
\end{equation}
where $\Omega$ is the solid angle. The integral constraint can be estimated 
(providing enough random points are used) with $IC=A_{w}B$ where $B$ is given 
by:

\begin{equation}
B = \frac{\sum RR(\theta)\theta^{-\beta}}{\sum RR(\theta)}.
\end{equation}

\noindent
The angular correlation function then becomes,

\begin{equation} 
w(\theta)=w_{obs}(\theta)+IC=A_{w}\theta^{-\beta}.
\end{equation}

\noindent
The best fitting parameters $A_{w}$ and $\beta$ can be found by finding the 
best ($\chi^{2}$ minimisation\footnote{The technique of $\chi^{2}$
minimisation assumes independent errors. In fact, the errors on $w(\theta)$ 
measurements are correlated. This complication is neglected here, but we note
that if $\chi^{2}$ is used to test goodness-of-fit, it is likely to be an 
underestimate.}) fit to the function
$w_{obs}(\theta)=A_{w}(\theta^{-\beta}-B)$, for different values of $\beta$, 
where $B$ is recalculated each time. The error 
on $A_{w}$ is computed as the value that gives $\Delta\chi^{2}=1$. 
 
Figure~\ref{fig:wtheta} shows the angular correlation function for the fainter 
sample of
66 LBG candidates covering an area where we are confident there is comparable 
completeness to $i'=24.5$. Some 90,000 random objects were used in this 
calculation. The angular correlation
function for the `bright' sample to $i'=24.0$ is shown in Figure
\ref{fig:wtheta2} based on 74 LBG candidates. Here, 250,000 random 
objects were used to calculate the correlation function. Since the clustering 
strength of faint galaxies is significantly less than that discussed below
for bright LBGs, faint galaxies in the same catalogue could also be used as
an effective sample of `random' objects. This provides a good test that there
is no remnant structure due to varying completeness in the $i'-$band data.
The galaxies to be used as `random' objects were selected to cover the same
magnitude range as the LBGs (i.e. $23.0<i'<24.5$ for the faint sample,
and $23.0<i'<24.0$ for the bright sample).  Similar correlation strengths to 
using a truly random sample were obtained. For the $i'<24.5$ sample a slightly 
higher value of $A_{w}=15.52$ was measured ($\chi^{2}_{{{\rm red}}}=1.01$). 
For $i'<24.0$ a value of $A_{w}=17.58$ was obtained 
($\chi^{2}_{{{\rm red}}}=0.89$).

There is increasing evidence that faint samples of $I$-band selected galaxies 
(and therefore galaxies with a higher median redshift) have a shallower 
slope, $\beta$, to the best fitting power law $w(\theta)=A_{w}\theta^{-\beta}$
than the often quoted $\beta=0.8$ seen in clustering studies of local 
galaxies \citep[e.g.][]{brainerd,postman,mccrack}. In addition, 
many measurements of LBG clustering also find a best fitting slope that is 
shallower than 0.8 \citep[e.g.][]{ouchi,p_and_g02}, although few of the 
measurements are accurate enough to test whether or not this is significantly 
different to the local Universe, and so results are usually quoted with a 
fit constrained to $\beta=0.8$. 

In the ODT data for $i'<24.5$ the best fitting parameters (with $\theta$ in 
arcsec) are consistent with this ($A_{w}=7.43^{+12.4}_{-5.0}$, $\beta=0.63^{+0.24}_{-0.23}$). However, there 
are not enough data to place a strong constraint on this slope. Therefore, to 
be consistent with other measures of LBG clustering in the
literature, only results for a constrained slope of $\beta=0.8$ are 
considered in what follows. However, we note that LBG correlation functions
with a shallower slope would have the general effect of \emph{increasing} 
estimates of $r_{0}$ (in this case by $\sim15\%$). This is something 
that will need to be looked into in more detail in the full ODT sample, and in 
other high quality data sets probing the high redshift Universe.  

Confining the slope of the angular correlation function
to 0.8, we obtain an amplitude $A_{w}$ of $15.39\pm4.3$ arcsec$^{0.8}$ for the 
faint $i'<24.5$ sample. For the brighter sample a best-fitting amplitude of 
$A_{w}=16.40\pm8.6$ arcsec$^{0.8}$ was calculated. The quoted errors in 
$A_{w}$ correspond to the error on the best fit to the power law 
$w(\theta)=A_{w}\theta^{-0.8}$. We make no attempt here to correct the 
correlation function for contamination, in line with similar 
studies with similar contamination estimates. However we note that, in the case
of maximum contamination our estimates of $A_{w}$ could be too low by 
$\sim20\%$\footnote{If the contaminants are unclustered then $A_{w}$ would be 
reduced by a factor $(1-f)^{2}$, where $f$ if the contamination fraction. In
the case where contaminants are clustered, and their clustering amplitude is
smaller than that for LBGs, the reduction in $A_{w}$ would be less.}
The effect on the deprojected $r_{0}$, as we discuss in the next 
section, will be smaller.

\subsection{Spatial Correlation Function}
\label{subsec:spatcorr}

Using our adopted redshift distribution, $N(z)$ (Section \ref{sec:Nz}), in
conjunction with the best fitting parameters to our angular
correlation functions, we estimate the corresponding correlation lengths for 
galaxies using the Limber deprojection \citep{peebles1980,efstathiou}. The
angular correlation function, $w(\theta)$, and spatial correlation function, 
$\xi(r)$, are related via:

\begin{equation}
A_{w} = r_{0}^{\gamma}C\frac{\int_{0}^{\infty} F(z)
D_{\theta}^{1-\gamma}(z) N(z)^{2} g(z)dz}
{\left[\int_{0}^{\infty}N(z)dz\right]^{2}},
\end{equation}
\noindent where C is a constant, 
\begin{equation}
C = \sqrt{\pi}\frac{\Gamma[(\gamma-1)/2]}{\Gamma({\gamma/2})}
\end{equation}

\noindent
$D_{\theta}(z)$ is the angular diameter distance, $g(z)$
is a function that describes the cosmology, 

\begin{equation}
g(z) =
\frac{H_{0}}
{c}[(1+z)^{2} (1+\Omega_{m}z+\Omega_{\Lambda}(1+z)^{-2}-1)^{1/2}], 
\end{equation}
and $F(z)$ is the redshift dependence of $\xi$(r). 

Since the selection function for LBGs is very narrow (especially as a function
of time), the function $F(z)$ can be removed from the integral. This can be
verified by using both linear and non-linear clustering evolution for the 
function $F(z)$ over our redshift distribution. We find this gives only 
negligible changes in $r_{0}$ \citep[c.f.][]{ouchi,p_and_g02}. The comoving 
$r_{0}$ at redshift $z$ is therefore given by: 

\begin{equation}
r_{0}(z)=r_{0}[F(z)]^{1/\gamma}.  
\end{equation}

Using our fiducial redshift distribution and the angular correlation function 
parameters derived in Section~\ref{subsec:angcor} we obtain a spatial
correlation length (at $z\sim4$) of $r_{0} = 11.4_{-1.9}^{+1.7} h^{-1}_{100}$ 
Mpc for the $i'<24.5$ sample.
For the $i'<24.0$ sample we obtain $r_{0} = 11.8_{-4.0}^{+3.1} h^{-1}_{100}$ 
Mpc (see Table \ref{tab:summary} for a summary of these results). These 
correlation lengths are considerably greater than other
measurements at $z\sim4$ \citep[e.g.][]{arnouts,arnouts02,ouchi} and the 
implications and context of this are discussed in Section \ref{sec:disc}.

In order to test the accuracy of these results we vary the $N(z)$ 
distribution introduced in Section~\ref{sec:Nz} over a reasonable range of 
values for $\bar{z}$ and $\sigma_{z}$ and also consider a simple top-hat 
redshift distribution (see Figure \ref{fig:Nz}). We find that the calculated 
$r_{0}$ is only weakly dependent on the $\bar{z}$ used, and is only affected 
at the $10\%$ level, even using a top-hat distribution over $3.5<z<4.5$. 
However, significant broadening of the selection function by increasing 
$\sigma_{z}$ to 0.4 would increase $r_{0}$ by up to 40\%. It should also be 
noted that narrower selection function would lead to smaller values of $r_{0}$.
A Gaussian selection function with $\sigma_{z}$ reduced to 0.1 would lead
to an $r_{0}$ that is $\sim$20\% smaller. If we 
consider the maximum final contamination given in Table \ref{tab:contam}, and 
assume that contaminants are unclustered (see Section \ref{subsec:angcor}), we
find that $r_{0}$ would increase by $15\%$. Note that our quoted results 
contain the error derived from the best fitting power law only. In line with 
other LBG studies, the effects of systematic uncertainties are not 
considered further.

\subsection{Linear Bias}
\label{subsec:bias}

The spatial correlation length $r_{0}$ can be used to calculate the linear
bias. 
Several definitions of linear bias appear in the literature; here we adopt
the definition based on the variation in mass over 8 $h^{-1}$ Mpc spheres 
($\sigma_{8}$) and bias is defined at our fiducial median redshift ($\bar{z}=4.0$).

Using our correlation function measurements, and assuming a
$\Lambda$CDM cosmology with an $n=1$ power spectrum of initial mass
fluctuations, normalised to $\sigma_8=0.9$, we use linear theory (neglecting 
any non-linear corrections which are expected to be small at $z\sim4$) to
determine the linear bias,

\begin{equation}
b={{\sigma_{{\rm g}}}\over{\sigma_{{\rm dm}}}},
\label{eq:bias}
\end{equation}

\noindent
all measured at a scale of $r=8$\,$h_{100}^{-1}$\,Mpc.  

In linear theory, the dark-matter variance, $\sigma_{dm}$, at redshift 
$z=0$ can be calculated for an arbitrary mass scale $M$, which is 
equivalent to a physical scale $R$, via:

\begin{equation}
R=\left(\frac{3}{4\pi}\,\frac{M}{\rho_{b}}\right)^{\frac{1}{3}},
\end{equation} 

\noindent
where the density, $\rho_{b}=\Omega_m\times(2.78\times10^{11})\,h_{100}^2$. 
The dark-matter variance, $\sigma_{dm}(z=0)$, can then be calculated for a 
given physical scale $R$, using the relation between mass-scale and variance.  
At redshift $z$, the dark-matter variance is given by: 

\begin{equation}
\sigma_{dm}(z)=\sigma_{dm}(z=0)\times D_{lin}(z), 
\end{equation}

\noindent
where $D_{lin}(z)$ is the linear growth factor at the redshift of interest; at 
redshift $4$, $D_{lin}(z=4)=0.25569$.  
The $\sigma_{\rm g}$ is calculated (assuming $\xi$ is described by a power law)
as:

\begin{equation}
\sigma_{\rm g}^2 = J_2 \left({{r_0}\over{r}}\right)^{\gamma},
\end{equation}

\noindent
where,

\begin{equation}
J_2 = {{72}\over{(3-\gamma)(4-\gamma)(6-\gamma)2^{\gamma}}},
\end{equation}

\noindent
\citep{peebles1980}. See \citet{somerville04} for further details and 
discussion.

For $i'<24.5$ we obtain $b=8.1_{-1.2}^{+1.1}$, and for $i'<24.0$, 
$b=8.4_{-2.6}^{+2.0}$. The clustering and bias results are summarised in 
Table \ref{tab:summary}.

\begin{table*}
\begin{center}
\begin{tabular}{cccccc} \\
\hline\\
Reference & Sample & $A_{w}$ & $r_{0}$ ($h_{100}^{-1}$\,Mpc) & $b$ & $n$ ($h_{100}^{3}$ Mpc$^{-3}$)\\ 
\hline
ODT (this paper) & $i'<24.0$ & 16.40 & $11.8_{-4.0}^{+3.1}$ & $8.4_{-2.6}^{+2.0}$& $(2.23\pm0.25) \times 10^{-5}$\\
ODT (this paper) & $i'<24.5$ & 15.39 & $11.4_{-1.9}^{+1.7}$ & $8.1_{-1.2}^{+1.1}$& $(8.86\pm0.49) \times 10^{-5}$\\
\citet{ouchi} & $i'<25.5$ & 0.97 & $3.2_{-1.2}^{+1.0}$ & $2.6_{-0.9}^{+0.7}$& $(1.71\pm0.07) \times 10^{-3}$\\
\citet{ouchi} & $i'<26.0$ & 0.71 & $2.7_{-0.6}^{+0.5}$ & $2.2_{-0.5}^{+0.4}$& $(3.72\pm0.11) \times 10^{-3}$\\
\citet{ouchi04} & $i'<24.8$ & 6.1 & $7.9_{-2.7}^{+2.1}$ & $5.3_{-1.7}^{+1.3}$ & $(2.2\pm0.6 ) \times 10^{-4}$\\ 
\citet{ouchi04} & $i'<25.3$ & 2.6 & $5.1_{-1.0}^{+1.1}$ & $3.5_{-0.7}^{+0.6}$ & $(9.2\pm1.10) \times 10^{-4}$\\ 
\citet{ouchi04} & $i'<26.0$ & 1.7 & $4.1_{-0.2}^{+0.2}$ & $2.9_{-0.1}^{+0.1}$ & $(4.9\pm0.30) \times 10^{-3}$\\ 
\hline\\
\end{tabular}
\caption{Summary of correlation function parameters, bias and number densities.
The bias is calculated as $b = \sigma_{{{\rm g}}} / \sigma_{{{\rm dm}}}$ using linear theory and a power spectrum via $n=1$ and 
$\sigma_{8}=0.9$.}
\label{tab:summary}
\end{center}	
\end{table*}

\section{Discussion}
\label{sec:disc} 

In hierarchical models of structure formation, structure forms by the
magnification of initial density fluctuations by gravitational instability. 
A key feature is that virialised structures (or dark matter haloes) should 
have clustering properties that differ from that of the overall mass 
distribution, with more massive haloes being more strongly clustered 
\citep[e.g.][]{kaiser}. If galaxies and clusters form when baryonic material 
falls into the potential wells of the dark matter haloes then a correlation
between galaxy (and hosting halo) mass and clustering strength (and therefore
bias) would be expected. 

\begin{figure}
\begin{center}
{\leavevmode \epsfxsize=8.0cm \epsfysize=8.0cm \epsfbox{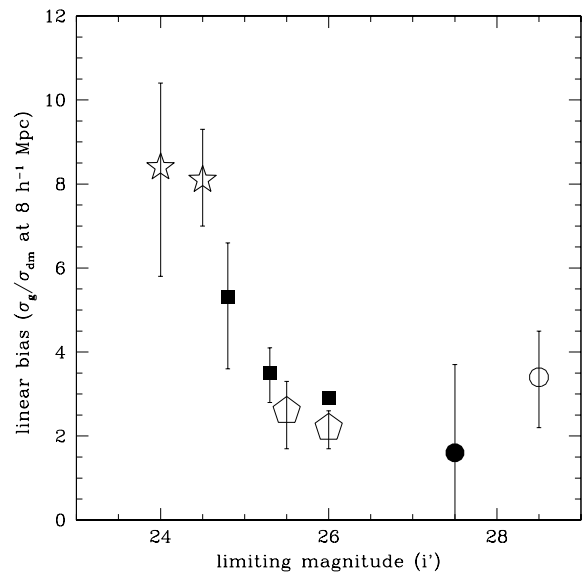}} 
\end{center}
\caption{The relationship between the limiting $i'-$band magnitude of a
survey and the measured bias (using the $\sigma_{8}$ definition as discussed in
the text). ODT data are represented by large open stars.
Data from \citet{ouchi} are shown as open pentagons. Filled squares correspond
to data from \citet{ouchi04}. The filled circle represents results from \citet{arnouts}, and the open circle shows results from \citet{arnouts02}.}
\label{fig:odt_siglim}
\end{figure}

An important measurement in studies of LBGs has been that of their clustering 
properties. The strong spatial clustering and surface densities 
exhibited by LBGs at $z\sim3$ appears to be consistent with 
biased galaxy formation, implying an association between LBGs and 
fairly massive dark matter haloes, even if there are several galaxies per dark
matter halo \citep[e.g.][]{giav98,g_and_d01,spf,wechsler01,bullock02}.
The measurement of the clustering properties of z$\sim$4 LBGs from
the ODT survey provides values of $r_{0}$ and $b$ that are significantly
larger than previous measurements of these parameters at this redshift
\citep{ouchi,arnouts}, and of U-band dropouts at $z\sim3$ \citep{giav98, 
arnouts, g_and_d01, p_and_g02}. However, the data presented here are from a 
sample that is much brighter than other measurements of the correlation
function at $z\sim4$. We are therefore able, by combining these shallower
wide-field data with deeper pencil-beam surveys such as the Hubble and Subaru 
deep fields, to compare with the luminosity dependent bias detected in the 
local Universe \citep{norberg} and at high redshifts 
\citep{g_and_d01,foucaud,ouchi04}, and test models of biased galaxy formation.   
Using our definition of bias (Equation \ref{eq:bias}) we calculate bias
parameters for the $z\sim4$ Subaru deep field clustering results of 
\citet{ouchi}. We obtain $b=2.6$ for $i'<25.5$ and $b=2.2$ for $i'<26.0$.
We also consider the more recent results of \citet{ouchi04} who present 
clustering results for several differently defined LBG samples. The three
samples using data up to a given magnitude limit (with $b=5.3$ for $i'<24.8$,
$b=3.5$ for $i'<25.3$, and $b=2.9$ for $i'<26.0$) are used here.
When these data are considered alongside the data presented in this paper, 
there appears to be a clear trend between the depth of a sample and LBG bias. 
At $z\sim4$, our faintest magnitude limit of $i'\approx24.5$
corresponds to a rest-frame absolute magnitude of $M_{\rm 1700}\approx
-20.5$ at $\lambda_{\rm rest}=1700$\,\AA\footnote{To calculate the
appropriate $k$-corrections, we have used the rest-frame UV SED of
\citet{shapley03}.}. Employing the $z\sim4$ luminosity function of
\citet{steidel99}, the faintest galaxies in our sample are then
$L_{\rm 1700}\approx1.5L_*$ (the brightest being $L_{\rm 1700}\approx6.1L_*$), 
whereas the \citet{ouchi} and \citet{arnouts,arnouts02} samples are, on the
whole, sub-$L_*$. The luminosity/bias results are summarised in Table 
\ref{tab:summary} and are plotted in Figure \ref{fig:odt_siglim}.

One interpretation of luminosity dependent bias could be a tight 
correlation between dark matter halo mass and star formation rate
\citep[i.e. UV luminosity;][]{g_and_d01,steidel98}. The observed 
$i'$-band magnitudes measured here correspond to a rest-frame
of $\lambda\sim1500$ \AA. At this wavelength we are measuring the
UV luminosity which is a good tracer of the star formation rate in the 
galaxy. We can therefore infer a link between halo
mass and star formation rate. However we emphasise that the 
rest-frame UV luminosity is more indicative of the \emph {instantaneous} star 
formation rate, rather than total underlying light (and therefore stellar
mass).

\subsection{The Masses and Environments of Bright $z\sim4$ LBGs}
\begin{figure}
\begin{center}
{\leavevmode \epsfxsize=8.0cm \epsfysize=8.0cm \epsfbox{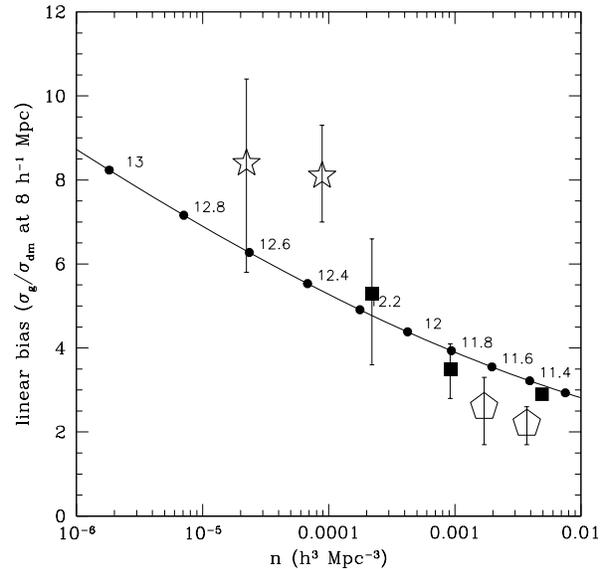}} 
\end{center}
\caption{The relationship between galaxy number density and the linear bias 
parameter, for a simple galaxy halo occupation function with $\alpha=0$ 
(i.e one galaxy per halo) is shown by the solid black line. The numbers 
marked on the line correspond to the dark matter halo masses hosting 
the observed galaxies in this model. Masses are written as log$_{10}$($M$) 
in units of $h^{-1}$ \msun. Open stars represent ODT data presented in 
this paper. Open pentagons represent the results of \citet{ouchi}. Filled
squares correspond to data from \citet{ouchi04}. See 
Table \ref{tab:summary} for a summary of these values.} 
\label{fig:odt_signum}
\end{figure}

In Figure \ref{fig:odt_signum} we plot bias versus comoving space
density for galaxies at $z\sim4$.  The observed space densities are
calculated in Section \ref{sec:sd}.  For the ODT bright sample of $i'<24.0$ 
and the faint sample of $i'<24.5$, we obtain bias values of 8.4 and 8.1,
respectively.  The corresponding space densities, calculated using the
same selection function used for the correlation function inversion,
are $2.23\times10^{-5}$ and $8.86\times10^{-5}\,h^3_{100}$Mpc$^{-3}$.
We also show points for the \citet{ouchi04} and \citet{ouchi} results, using 
their estimated effective volume and the reported numbers of LBGs for each
limiting magnitude (with estimated uncertainties). 

The line in Figure \ref{fig:odt_signum} depicts the correspondence between
linear bias and comoving space density for dark matter halos at $z=4$
(based on Sheth \& Tormen 1999; see also Somerville et al. 2004).
Smaller space densities correspond to more rare overdensities, and therefore
to higher minimum dark matter halo masses (as shown).  This is the
relation that $\Lambda$CDM predicts that galaxies would follow, if
there were only one galaxy in each dark matter halo, which is a useful
simplification for checking trends, but not how galaxies tend to exist.
In cases where there may be varying numbers of galaxies per halo,
particularly likely if these are starbursting galaxies, the `halo
occupation' formalism is a useful parametrisation. Here, above some
threshold minimum mass, we assume there is a function 
$N(M>M_{min} | \alpha,M_1)$ that describes both the internal mass-function 
slope ($\alpha$), and the mass of a dark matter halo that typically hosts one 
galaxy ($M_1$; therefore, there may statistically be $M>M_{min}$ halos that
have no resident galaxies).  This prescription is found to describe
galaxies at both $z\sim0$ very well \citep*{marinoni,bosch03,hof_2df} and
$z\sim3$ \citep[e.g.][]{wechsler01,bullock02,kravtsov}.

A detailed exploration of the occupation statistics of LBGs implied by
our measurements is beyond the scope of this paper. It is useful,
however, to turn to the one-galaxy-per-halo curve, and note two items.
First, the data span a large dynamic range of space densities, and yet
follow the same general trend as the haloes.  This suggests that the
generic $b$/$n$ correlation predicted by $\Lambda$CDM is reflected in the data.
Second, the mass-scale of the ODT LBG sample, $M_{min}\sim {\rm
few}\times 10^{12} h^{-1}$\,M$_{\odot}$, begs the question whether
these naturally connect with later galaxy populations known to be in
similarly massive environments.  In \citet{ms02}, the possible
clustering evolution between massive galaxy populations at $z\sim0$,
$1.2$, and $3$ were explored in the context of simple
$\Lambda$CDM-motivated models.  The $z\sim3$ galaxy population
characteristics were drawn from the literature of largely-sub-$L_*$
samples, and were shown to not have a natural connection with the
lower-redshift ($z\sim1.2$) populations, which are plausibly connected with 
the most massive (elliptical-galaxy-type) environments today.  The
$z\sim1.2$  (ERO) population's linear bias and comoving space density, when
extrapolated to $z\sim4$ \citep[under the `constant minimum mass' model;][]{ms02} are expected to have $b\approx8$ and
$n\approx10^{-5}h^3_{100}$\,Mpc$^{-3}$ -- values remarkably close to
those measured in the super-$L_*$ ODT LBG sample.

\subsection{Significance of Results}

Thus far the quoted errors on our measured $r_{0}$ and bias have only included
`random' errors based on the best fit to the angular correlation function.
In Section \ref{sec:Nz}, we derived our redshift distribution based on an
analysis of theoretical colours of high redshift galaxies. We also considered
a range of possible median redshifts and found that this led to a $\pm$10\% 
uncertainty on the derived value for $r_{0}$. Combining this error in 
quadrature with the measured statistical errors leads to 
$r_{0}=11.4^{+2.0}_{-2.2}$ and $r_{0}=11.8^{+3.3}_{-4.1}$. This is a difference from the results of \citet{ouchi} at the 3.7-$\sigma$ level for the faint 
sample and 2.1-$\sigma$ for the bright sample. 

In Section \ref{subsec:spatcorr} the possibility of broader and narrower 
redshift distributions than 
the one we assume was also discussed, leading to possible increases in $r_{0}$
of up to 40\%, and decreases in $r_{0}$ of up to 20\%. Using a
$+$40\%/$-$20\% uncertainty for the redshift distribution gives
values of $r_{0}=11.4^{+4.9}_{-3.0}$ and $r_{0}=11.8^{+5.6}_{-4.6}$ for the 
faint and bright samples respectively. This would be enough to reduce the 
difference from the \citet{ouchi} results to the  2.7-$\sigma$ and 
1.9-$\sigma$ level. 

We also noted in Section \ref{subsec:angcor} that the effect on 
contamination on our LBG sample is likely to dilute the correlation function, 
and any correction that might be applied would lead to an \emph{increase} in 
$r_{0}$, further strengthening any relationship between luminosity and bias.

\subsubsection{Cosmic Variance}

Since the LBG population is known to be strongly clustered, it is clear that 
the number density of objects will vary from field to field. It is therefore 
important that a large enough volume of sky is surveyed so that this `cosmic
variance' does not significantly bias the measured number density of objects
relative to the overall cosmic average. In order to estimate the size of such 
an effect on this data set we use the results of \citet{somerville04} 
who compute the variance of dark matter from linear theory as a function of 
redshift.

Based on the estimated effective volume at $z\sim4$ covered in this survey 
($3\times10^{4}$ Mpc$^{3}$, assuming $h=0.7$), the variance in the mean 
dark matter density is estimated to be 15\% 
\citep[see Figure 3b in][]{somerville04}.
How cosmic variance relates to an uncertainty in number density of a 
population depends on how biased that population is relative to the overall
dark matter distribution.
Assuming the bias measured in Section \ref{subsec:bias} is correct, then the 
uncertainty in the number density of LBGs due to cosmic variance can be 
estimated as $\sigma_{{{\rm cv}}}=8.1\times0.15=1.2$.  
Therefore, in Figure 13 there is an additional uncertainty in the estimated 
space density due to the sample size (volume) and cosmic variance associated 
with this. 

Could fluctuations in number density caused by cosmic variance produce the 
observed difference in clustering strength between the ODT and fainter 
samples? This is difficult to test since estimates of the size of cosmic 
variance effects from linear theory require prior knowledge of the 
relationship between the galaxy population being studied and the underlying 
dark matter distribution (i.e. bias). 
However, if we start with the null hypothesis that the intrinsic bias of the 
ODT LBG sample is actually the same as that measured by \citet{ouchi}, then 
the uncertainty in number density due to cosmic variance, 
$\sigma_{{{\rm cv}}}=2.6\times0.15=0.39$ \citep[following][]{somerville04}.
If we have \emph{underestimated} the intrinsic number density due to 
cosmic variance, and assume that $\xi\propto\,1/(n^{2})$ , then the `true' 
number density could indeed be larger, and the correlation length (and hence 
bias) could be smaller\footnote{This approximation ($\xi\propto\,1/(n^{2})$) 
assumes that Gaussian statistics apply, which is not the case here. A full 
calculation of the effects of cosmic variance requires knowledge of higher
order clustering statistics. Simulations by \citet{foucaud} suggest that for a 
similar bright sample but at $z\sim3$, the size of the cosmic error is likely 
to be smaller than the measured Poisson errors. However, this rough 
calculation provides a useful measure of the potential size of the 
`cosmic error'.}. Taking $\sigma_{{{\rm cv}}}=0.39$, and assuming a 
2-$\sigma$ fluctuation, implies an upper limit on the number density of 1.58 
$\times10^{-4}\,h^{3}_{100}$ Mpc$^{-3}$. 
If this corresponds to the \citet{ouchi} correlation length of 
$r_{0}=3.2\,h^{-1}_{100}$ Mpc, then our measured value for the number density 
(8.86$\times10^{-5}\,h^{3}_{100}$ Mpc$^{-3}$) would scale $r_{0}$ to 
$6.1\,h^{-1}_{100}$ Mpc which is inconsistent with our measured values (in fact
a 5.5-$\sigma$ fluctuation in number density is required to scale the 
\citet{ouchi} value to $11.4\,h^{-1}_{100}$ Mpc). 
Variations in number density due to cosmic variance are too small to 
explain the difference in clustering strength between the ODT and fainter 
samples. However, it is clear that wide field surveys at least as large as the 
ODT are required to measure the clustering properties of the brightest objects.
We also note that the independent results of \citet{ouchi04} are consistent 
with the trend of increasing clustering strength with brighter magnitudes at
$z\sim4$.

\subsection{Small Scale Clustering and Close Pairs}

In addition to a luminosity dependent bias, \citet{p_and_g02} provide evidence 
from a fairly bright sample of $z\sim3$ LBGs that there is also a significant
scale-dependent bias. In their data set there appears to be a lack of power
in the angular correlation function (and therefore a lower bias) on scales 
less than 30 arcsec. \citet{bullock02} also demonstrate (where redshifts are
available) how close pair statistics can be used within the halo occupation 
function formalism to place constraints on different models for the occupation 
function. \citet{p_and_g02} interpret the `break' in their correlation function
as evidence that the dark matter haloes have a fairly large size and mass, 
which is more consistent with one-galaxy-per-halo occupation functions. 
However it should be noted that our analysis (and that of most
other LBG clustering studies) has assumed a simple
power law for the correlation function with $\gamma=1.8$. In addition, the 
results of \citet{kravtsov} (which come from $N$-body simulations) indicate 
that the correlation function may actually steepen on the smallest scales, 
(which in turn would lead to over-estimates of $r_{0}$ for LBGs). The models
of \citet{hamana} indicate that the observed angular correlation function
is well fitted by a two-component power law with a steep component dominating 
on small scales, although \citet{ouchi04} fit the same data with single 
component model with a relatively steep ($\beta=0.9$) slope.

In the ODT data presented here we find no close pairs on scales less than
10 arcsec in the $i'<24.5$ sample and none less than 40 arcsec in the 
brighter $i'<24.0$ sample. Unfortunately, since only one or less pairs would be
required to fit the best-fitting power law on these scales, it is not possible
to determine whether or not this constitutes evidence to support the results
of \citet{p_and_g02}, or indeed to determine whether a steeper slope is more 
appropriate on small scales. However an analysis of the full ODT survey should 
yield a good measurement of the small-scale clustering of bright LBGs, and help
assess the significance of any lack of power on small-scales in the angular
correlation function.

\section{Conclusions} 
\label{sec:conc}

The main results presented here can be summarised as follows:

\begin{itemize}

\item The clustering of LBGs at $z\sim4$ is well approximated by a power 
law $w(\theta)=A_{w}\theta^{-\beta}$. 
Fixing the slope to that of local galaxies ($\beta=0.8$), provides a good 
fit to the data with $A_{w}=15.39$ arcsec$^{0.8}$ the best amplitude for 
$i'<24.5$ and $A_{w}=16.40$ arcsec$^{0.8}$ for $i'<24.0$. However we note that 
there is evidence to suggest that $\beta=0.8$ may not be a valid assumption to 
make at high redshift, and a shallower slope may be more appropriate. 

\item Using a reasonable fiducial redshift distribution to characterise the 
selection function for LBGs, the angular correlation function can be 
deprojected to obtain the spatial correlation length, $r_{0}$. For our 
fainter sample ($i'<24.5$) we obtain $r_{0}=11.4_{-1.9}^{+1.7}h^{-1}_{100}$ 
Mpc. Using a slightly brighter sample ($i'<24.0$) we obtain a similar 
correlation length of $r_{0} = 11.8_{-4.0}^{+3.1} h^{-1}_{100}$ Mpc. 
Comparing these results to the clustering properties of dark matter, we obtain 
linear bias values of $8.1_{-1.2}^{+1.1}$ and $8.4_{-2.6}^{+2.0}$ respectively.

\item When compared with fainter surveys the bias and correlation lengths
seen in the ODT are clearly significantly larger. We interpret this as 
evidence for a luminosity dependent bias for LBGs, as predicted by some
semi-analytic models. The ODT bias values are shown to be consistent
with a simple model using a galaxy occupation function describing one 
observable galaxy per dark matter halo. Bright (super$-L_{*}$) LBGs seem to 
be more biased tracers of mass than fainter (sub$-L_{*}$) ones, suggesting a 
relationship between halo mass and instantaneous star formation rate.

\item The population of sub$-L_{*}$ LBGs are unlikely to be the 
progenitors of massive galaxies at $z\sim1$ and $z\sim0$. However, the 
extremely bright, super$-L_{*}$ population presented in this paper have 
biases, number densities and halo masses ($M_{min}\sim {\rm
few}\times 10^{12} h^{-1}$\,M$_{\odot}$) that are consistent, in a simple 
model, with them being potential progenitors of the most massive galaxies.

\end{itemize}

The data presented here are only a small sample of the full ODT survey. With
an extended sample containing $>1000$ LBGs over a much wider area
it should be possible to place stronger constraints on the clustering 
properties of this bright sample of high redshift galaxies, and address issues 
such as their scale dependent bias. 
In addition, other surveys currently underway such as the NDWFS \citep{NDWFS}, 
GOODS \citep{GOODS}, the VIRMOS-VLT deep survey \citep{virmos}, the CFHT 
Legacy survey\footnote{http$://$www.cfht.hawaii.edu/Science/CFHTLS}, SXDS
\citep{SXDS}, and COSMOS\footnote{http$://$www.astro.caltech.edu/$\sim$cosmos}, should provide good measurements of LBG clustering probing (when combined) a 
large dynamic range in luminosity and redshift. This 
should allow tighter constraints on parameters such as the galaxy occupation 
function, and the scale and luminosity dependent bias for LBGs; thus providing 
a better understanding of the relationship between LBGs, other high-z 
populations, and galaxies today.

\section*{Acknowledgements}
The Isaac Newton Telescope is operated on the island of La Palma by the 
Isaac Newton Group in the Spanish Observatorio del Roque de los 
Muchachos of the Instituto de Astrofisica de Canarias.
PDA, CEH, and CAB acknowledge the support of PPARC Studentships. ECM thanks 
the C.K. Marr Educational Trust.
This work was supported by the PPARC Rolling Grant PPA/G/O/2001/00017 at the
University of Oxford.
LAM acknowledges support from the {\it SIRTF} Legacy Science Program, 
provided through an award issued by the Jet Propulsion Laboratory, 
California Institute of Technology under NASA contract 1407. Thanks to both
anonymous referees whose comments helped to improve this paper.

\bibliographystyle{mn}
\bibliography{thesis_references}

\end{document}